\documentclass[conference]{IEEEtran}
\IEEEoverridecommandlockouts
\usepackage{cite}
\usepackage{amsmath,amssymb,amsfonts}
\usepackage{algorithmic}
\usepackage{graphicx}
\usepackage{textcomp}
\usepackage{xcolor}
\usepackage{eso-pic}
\usepackage{multirow}

\def\BibTeX{{\rm B\kern-.05em{\sc i\kern-.025em b}\kern-.08em
    T\kern-.1667em\lower.7ex\hbox{E}\kern-.125emX}}
\begin{document}
\newcommand\AtPageUpperMyright[1]{\AtPageUpperLeft{
 \put(\LenToUnit{0.28\paperwidth},\LenToUnit{-1cm}){
     \parbox{0.75\textwidth}{\raggedleft\fontsize{9}{11}\selectfont #1}}
 }}
\newcommand{\conf}[1]{
\AddToShipoutPictureBG*{
\AtPageUpperMyright{#1}
}
}

\title{Evaluation of Software Quality Through Refactoring Commit-Impact Analysis\\
\conf{2023 International Conference on Next-Generation Computing, IoT \& Machine Learning (NCIM 2023) 16-17 June 2023, Gazipur-1707, Bangladesh}
 {\footnotesize 
 }
\thanks{979-8-3503-1600-1/23/$31.00 ©2023 IEEE$}
}


\author{
\IEEEauthorblockN{
Md Safayat Hossen, Sheikh Moonwara Anjum Monisha, Shadid Intisar Quasem
}
\IEEEauthorblockA{
Department of Computer Science and Engineering, 
Islamic University of Technology, Bangladesh\\
safayathossen@iut-dhaka.edu, moonwaraanjum@iut-dhaka.edu, shadidintisar@iut-dhaka.edu
}
}

\maketitle

\begin{abstract}

The aim of this study is to conduct an empirical analysis of the correlation between code modifications and software quality. This will be achieved by examining the effects of changes made before and after each refactoring commit, using software metrics such as size, complexity, code smells, vulnerability, and bug detection. The objective is to conduct static code analysis on open-source software from GitHub repositories using various repository mining and static code analysis tools. Our approach involves formulating research-based questions and utilizing the specified tools to obtain results that will enable us to resolve the case study. The obtained results have been summarised and further analyzed using tables and graphs to achieve the objective.
\end{abstract}


\section{Introduction}

Software development is an ongoing process. To increase the system's overall efficiency and effectiveness, multiple functionalities must be integrated and maintenance processes must be continuously improved. Usually Software tends to get bigger and more complex as it develops. Evolutionary processes may deviate from the accepted coding standards. When building a module, it can be difficult to keep up with coding rules and standards. Refactoring can be applied to them at a later time, though.

Refactoring a codebase is a software development practice that generally results in improved software quality. Refactoring, as defined by Fowler \cite{b10}, is the process of altering the source code of a software project in order to improve its internal structure. Improving measurable internal quality attributes such as cohesion, coupling, complexity, inheritance, and size is a feasible method for enhancing the structural quality of code \cite{b11}. 

Numerous version control systems such as Git and CVS have aided the collaborative software development process. These systems have proven to be convenient by enabling the tracking of each change, also known as a commit. The historical record facilitates code sharing among developers and enables them to monitor modifications made to software systems during both the development and maintenance phases. Furthermore, it serves as a valuable data source for comprehending the software development process\cite{b12}.Commits in software development provide a record of the modifications made to the codebase, including the identity of the individual who made the changes and the timestamp of the modifications. There is a plethora of information available on the evolution of software.

An analysis of the impact of commits that have been made. Refactoring is a standard software development practice that involves evaluating the software's quality before and after each refactoring commit. The aforementioned approach holds promise in providing substantial insights into the growth of software and the impact of specific alterations on its overall quality. This article explores the impact of individual refactoring contributions on the source code, the correlation between source code modifications and software quality metrics, and the effectiveness of a specific metric as a software quality indicator. Our focus will be directed towards the Refactoring Commit Impact Analysis. The effects of changes made before and after each refactoring commit are examined by utilizing software metrics such as size, complexity, code smells, vulnerability, and bug detection. This is achieved by conducting static code analysis on open-source software from GitHub repositories using various repository mining and static code analysis tools. To represent our research, we have formulated four research questions.

\textbf{Relevant research-based questions}
\begin{itemize}
\item RQ1: Do refactoring commits have any impact on quality metrics?
\item RQ2: What is the impact of refactoring on code smells?
\item RQ3: Does the refactoring commit impact the same way on novice and mature projects?
\item RQ4: How does refactoring commit impact projects over time?
\end{itemize}

\section{Related Work}

Throughout the software lifecycle, refactoring is a crucial factor in enhancing the internal quality of software. A large amount of studies have been performed to analyze the impact of refactoring operations on software quality. Many researchers also worked to gain insight into the evolution of software systems through repository mining. These studies greatly influenced us to perform our case study on this subject. 

Pooyan Behnamghader \cite{b4} conducted an empirical study to understand the software quality evolution through commit impact analysis. They explored whether each commit has an impact on the source code, examined how source code changes affect software quality metrics, and studied the effectiveness of using a certain metric as a software quality indicator. Alexander Chavez\cite{b5} performs a study to investigate how refactoring operations affect five well-known internal quality attributes: cohesion, coupling, complexity, inheritance, and size. Oscar Chaparro \cite{b6} presents RIPE (Refactoring Impact PrEdiction), a technique that estimates the impact of refactoring operations on source code quality metrics. RIPE supports 12 refactoring operations and 11 metrics and it can be used together with any refactoring recommendation tool. Oumayma Hamdi\cite{b7} conducted an empirical study on the influence of refactoring on software quality in the context of mobile apps, applying the difference-in-differences (DiD) model to analyse the impacts of 10 frequently used refactoring operations on 10 well-known quality metrics in Android apps. They discovered that, several metric values improved significantly as a result of some refactoring types; refactoring often has little to no effect on the metrics, however, one metric (LCOM) actually degrades as a consequence of refactoring, and other than the design size metrics, the results from the nonrefactoring commits show no notable distinction. Islem Saidani\cite{b8} explores into how the quality-driven continuous integration (CI) process could impact how developers refactoring approach. According to the results of the analysis of the changes using Multiple Regression Analysis (MRA), software developers require more specialized refactoring tools in the context of CI to better maintain and improve their software systems. Alexander Ch´avez\cite{b9} studied how refactoring operations affect five well-known internal quality attributes. They analyzed different refactoring tactics and found that when they examined several refactoring techniques, they discovered that in 65\% of the cases, the associated internal quality attributes are enhanced, and in the other 35\% of operations, the quality attributes are left unchanged. While reaching specific aims, 55\% of refactoring operations enhance internal quality attributes.

\section{Tools}

We used a static code analyzer for source code analysis and a repository mining tool to detect the refactored commit from the repository. Tools for static code analysis provide quality metrics. These tools are capable of analyzing software without running its source code. These tools reveal software defects, which can be useful for identifying early-stage errors made by developers that cannot be traced to the compiler because they are primarily design flaws. We utilized three tools for static code analysis and two for repository mining.

\subsubsection*{\textbf{Refactoring Miner 2.1}}
A library or API written in Java that can detect refactorings applied in the history of a Java project \cite{b1}. It can detect 81 types of refactorings, such as Extract Method, Inline Method, Rename Method, Move Method, etc. Though it was built for Java repository mining, now Python and Kotlin-supported versions are also available. It can be run from the command line and also from a Chrome extension. Gradlew, a wrapper package, is required to run this tool on Java repositories.

\subsubsection*{\textbf{PYREF}}
PYREF is a software tool designed for the purpose of extracting refactoring operations from Python projects. It is a useful resource for developers seeking to analyse and improve the quality of their code. The tool is referenced in \cite{b2}. The tool is based on the approach used in REFACTORINGMINER \cite{b1}, from which it draws inspiration. The primary methodology employed in this tool closely resembles that of REFACTORINGMINER. It is a software tool that accepts a Git repository containing a Python project as its input. It then generates a comprehensive list of refactoring operations that have been executed within the project. The software system is capable of detecting various method-level refactoring operations, including Rename Method, AddParameter, RenameParameter, ExtractMethod, InlineMethod, MoveMethod, PullUpMethod, and PushDownMethod.

\subsubsection*{\textbf{PMD (Programming Mistake Detector)}}
The tool functions as a source code analyzer, detecting typical programming errors such as redundant code, superfluous if statements, empty catch blocks, unused variables, and other similar issues. The software provides support for various programming languages including Java, JavaScript, PLSQL, Apache Velocity, XML, XSL, and more. Furthermore, it encompasses CPD, which stands for copy-paste detector. The CPD tool is capable of detecting duplicated code in a variety of programming languages, including Java, C, C++, C\#, PHP, Ruby, Fortran, JavaScript, PLSQL, Apache Velocity, Scala, Objective C, Matlab, Python, Go, Swift, and others. The software incorporates multiple sets of rules for every programming language it is compatible with. The Java language is supported by over twenty rule sets. Rule sets may have overlapping rules, which implies that a rule can be a part of more than one rule set.

\subsubsection*{\textbf{Source Monitor}}
The tool is designed to measure source-code metrics and offers the capability to store measurement results persistently. Additionally, it allows for a historical comparison of the saved results. The tool collects cyclomatic complexity statistics and various other metrics for multiple programming languages. In addition to its primary function, Source Monitor provides information on eleven other metrics that can be of use. These metrics include the total count of lines in code files, the count of classes, the average number of methods per class, and the count of statements per method. The aforementioned outcome establishes a set of fundamental measurements that can be utilised to evaluate every subsequent analysis iteration, in order to oversee the comprehensive intricacy of your project. The "Complexity versus Features" curve can be a useful tool to comprehend the impact of adding new features to your software. The Source Monitor tool quantifies and evaluates the quality of source code written in programming languages such as C++, C, C\#, VB.NET, Java, Delphi, Visual Basic (VB6), and HTML.

 \subsubsection*{\textbf{SonarQube }}
The Code Quality Assurance tool gathers and evaluates source code, generating reports on the quality of code within your project. Although the various analyzers offer comparable functionalities, they can be incorporated into diverse development environments. The tool integrates static and dynamic analysis techniques and facilitates ongoing quality measurement. SonarQube offers a range of software quality metrics, including complexity, NCLOC (Non-Commented Line Of Code), and the number of classes. The software offers significant metrics, including bugs, vulnerabilities, and code smells.
 
\section{Methodology}
\subsection{Project Selection}\label{AA}
To find the answers to our research questions, we analyzed four open-source software projects from GitHub. two of them are in Java language and the remaining two are in Python. The projects were searched by applying different filters e.g programming language, no of commits, no of issues, no of contributors, etc. using a GitHub repository search tool named SEART\cite{b3}.

For the first Java project, we chose a small-scale project named IutAssistance developed by novice developers. The project has a total of 69 commits and among those 24 are refactor commits. The second Java project, Alligator has 198 commits and 113 are refactoring commits. The GitHub repository has 289 stars.

The first Python project is you-tube-search which has 168 commits and 18 refactor commits. The GitHub repository has 308 stars. The second project is named mini-shop-server which has 506 commits and 90 refactor commits. The repository has 500 stars.

\subsection{Refactor commit selection:}\label{AA}
Our goal was to observe the change in software metrics after refactoring commits. After selecting our object systems, the first task was to collect refactoring commits from each of the projects. For this purpose, we used a popular refactoring miner tool for Java and Python. For Java repositories, we used RefactoringMiner 2.1. For Repository Minor 2.1, we first set up this tool on our machine. JDK 14 and gradlew 6.3 is required to set up this tool. Then we clone the Java project repositories inside the tools bin directory, then run each project’s commit extracting commands (ex. RefactoringMiner -a, [which detects all refactorings at \texttt{<branch>}for \texttt{<git-repo-folder>}. If \texttt{<branch>}is not specified, commits from all branches are analyzed.] ). In Java project 1, 24 refactoring commits out of 69 commits were found, and in Java project 2, 113 refactoring commits out of 198 commits were found. We needed to examine the commit before the refactoring commit to check the change in quality metrics after refactoring. We couldn't run the tool on all of the commits from the projects because of our limitations and time constraints. So, we had to manually select 16 commits (8 pairs) and 18 commits (9 pairs) from Java project 1 and Java project 2, respectively. To collect the refactoring commits from the Python project, we used PyRef. First, the project needed to be cloned using the command-line interface. After cloning, the command for extracting the refactor commits is run. The output is a list of refactoring operations performed in the project. We get a total of 13 refactoring commits from Python project 1 and 80 refactoring commits from Python project 2.

As we wanted to check the change in quality metrics after refactoring, we also needed to examine the commit before the refactoring commit. So we manually checked the refactor commits and the commit before it. Therefore, we examined a total of 26 commits from python project 1. We created separate project files for each 26(13-pair) revisions and ran the static analysis tools on each of the revision files. We randomly chose 18 pairs of commits and ran sonarqube on each of these revision files.

\subsection{Tool application}
To observe the impact of a refactoring commit, we used a static code analyzer named PMD. It is an extensible, cross-language code analyzer. We can use this tool in various languages. But in this research, we applied this tool only to Java. We used the default ruleset for PMD, where we calculated Best practices, Code style, Design, Documentation, and Error-prone.

To run PMD, we need to have Eclipse installed on our device, and we need to add the PMD plugin to Eclipse. To check the impact of a refactoring commit, we first have to upload the code of the previous version of the refactoring commit. We need to apply PMD to this and get the desired value. Then we have to upload the source code of the refactored commit and apply PMD to it to get the desired value. Then we will do a comparison with the achieved values.

We also used Source Monitor for Java projects. We added the source files by extension and all subdirectories of the projects. We applied some filters, e.g., the use of a modified complexity metric, ignoring continuous header and footer comments, and excluding blank lines on the search. The project file storage format was set to by default. After selecting a checkpoint name, the popup window containing all the files we wanted to work with was previewed. Files can be added or removed from the checkpoint as per the requirements. We got 15 different types of metrics results after finishing execution.

We executed Sonarqube on Python projects. First, we set up the Sonarqube server to run it on the browser. We ran our projects locally, and for that, we had to configure new projects for each revision file. Sonar-scanner for Python was used to scan the Python files and generate the results in a web browser. Sonarqube gives results for various metrics, e.g., vulnerabilities, bugs, duplications, security hotspots, code smells, and debt. But for our selected projects, we got the results of two metrics: code smell and debt.

\begin{table*}[!h]
\caption{ Metrics used in the study}
\resizebox{\linewidth}{!}{
\begin{tabular}{|c|c|c|c|}

\hline
Language & Tool & Metrics & Description\\
\hline
\multirow{14}{*}\centering Java  & \multirow{8}{4mm} \centering PMD & Best practice & Violation of Best Practise \\
\cline{3-4}
  & & Code style & Violation of Code style \\
\cline{3-4}
 & & Design & Violation of Design \\
\cline{3-4}
 & & Documentation & Violation of Documentation \\
 \cline{3-4}
 & & Error prone & Violation of Error prone \\
  \cline{2-4}
  & \multirow{10}{1mm} \centering Source monitor & Branches Separately counted & Statements that break the sequential execution \\
\cline{3-4}
 & & Classes  & Number of classes and interfaces in a version \\
 \cline{3-4}
  & & Methods/Class & Number of methods/functions in a class\\
 \cline{3-4}
  & & Avg Stmts/Method  & Avg number of statement in a method\\
 \cline{3-4}
  & & Avg Depth & Average nested block depth weighted by depth.\\
 \cline{3-4}
  & &  Avg Complexity  & \multirow{2}{1mm} \centering   Measure of the overall complexity measured for each method\\
  & &   &  (and, if present, each function)
in a file or checkpoint. \\
   \cline{3-4}
  & & LOC  & Line of code.\\
   \cline{3-4}
  & & Statements & Executable line of code\\
   \cline{3-4}
  & & Files  & Number of code/script files\\
 \hline
\multirow{10}{*}\centering Python  & \multirow{5}{4mm} \centering  Sonarqube &  Debt Code smell & Technical effort to fix the code smell\\
 
\hline
\end{tabular}
}
  \label{tab: Semantic Features of Question}
\end{table*}

\subsection{Metrics selection}
 We chose some basic and major metrics to analyze the projects. Basic metrics measure the size of a software system. For the basic group, we selected lines of code, statements, and classes. Major metrics are avg complexity, error-prone, code smell, bugs, etc. 
 
 We have calculated code smell, debt, and bugs using Sonarqube, as it is the most widely used popular tool to calculate major software metrics. Code Smells (SM) are pieces of code that make the system hard to maintain. We have calculated the number of code smells to see the impact of refactoring on the number of code smells after committing a refactoring change. We got the value of debt which is the effort to fix the code smells, it is represented in time (day, hour, minute). A bug is a number of codes that is demonstrably wrong or highly likely to yield unexpected behavior.
 
\subsection{Data Collection}
We executed PMD and SourceMonitor on java projects and got the result of a total of 21 metrics. We executed Sonarqube on python projects and got the result of two metrics: code smell and technical debt. Then we stored the results in an excel sheet for further analysis. The results of metrics achieved from running the tools on different revision files are shown in \tablename~\ref{tab2} , \tablename~\ref{tab3}, \tablename~\ref{tab4} and \tablename~\ref{tab5}.

\begin{table*}[!h]
\centering
\caption{Data collected from Java Project 1}
\begin{center}
\resizebox{\linewidth}{!} {
\begin{tabular}{|c|c|c|c|c|c|c|c|c|c|c|c|}
\hline
\multicolumn{7}{|c|}{\textbf{Source Monitor On Java Project 1}}& \multicolumn{5}{|c|}{\textbf{PMD on Java Project 1}} \\
\hline
Commit & Type & \%Branches & Class & Methods/Class & Avg Statement/Method & Avg Depth & Avg Complexity & Best Practices & Code Style & Design Documentation & Error Prone \\
\hline
Before &  3.6 & 38 & 1.87 & 4.72  & 1.23 & 1.85 & 94 & 284 & 196 & 118 & 136 \\
\hline
After & 3.4 & 42 & 1.86 & 4.83 & 1.27 & 1.79 & 125 & 317 & 222 & 132 & 165 \\
\hline
Before & 3.3 & 49 & 2.24 & 4.57 & 1.38 & 1.7 & 366 & 1132 & 470 & 352 & 408 \\
\hline
After & 3.3 & 49 & 2.24 & 4.56 & 1.37 & 1.7 & 364 & 1136 & 470 & 354 & 396 \\
\hline
Before & 2.7 & 65 & 1.94 & 5.1 & 1.37  & 1.79 & 416 & 1498 & 592 & 436 & 502\\
\hline
After & 2.7 & 67 & 1.91 & 5.15 & 1.38 & 1.8 & 416 & 1534 & 596 & 440 & 508\\ 
\hline
Before & 2.7 & 67 & 1.91 & 5.15 & 1.38 & 1.8 & 204 & 582 & 364 & 204 & 299 \\
\hline 
After & 3.5 & 76 & 2.11 & 5.49 & 1.49 & 1.88 & 242 & 761 & 506 & 249 & 358\\
\hline
Before & 3.8 & 176 & 3.01 & 3.87 & 1.45 & 1.67 & 882 & 5202 & 1814 & 1964 & 1358 \\
\hline
After & 3.8 & 177 & 2.98 & 3.88 & 1.45 & 1.67 & 888 & 5192 & 1814 & 1954 & 1356 \\
\hline 
Before & 4 & 159 & 2.82 & 4.23 & 1.48 & 1.75 & 838 & 4752 & 1716 & 1726 & 1264 \\
\hline 
After & 4 & 158 & 2.83 & 4.23  & 1.48 & 1.75 & 826 & 4732 & 1714 & 1718 & 1256 \\
\hline Before & 4 & 155 & 2.88 & 4.22 & 1.49 & 1.75 & 425 & 1889 & 1045 & 806 & 715 \\
\hline After & 4 & 148 & 2.86 & 4.37 & 1.5 & 1.78 & 415 & 1855 & 1035 & 776 & 705 \\
\hline
Before & 4.4 & 127 & 2.87 & 4.82 & 1.56 & 1.85 & 724 & 4302 & 1606 & 1422 & 1142\\
\hline
After & 4.4 & 124 & 2.85 & 4.84 & 1.54 & 1.83 & 702 & 4226 & 1568 & 1382 & 1120\\
\hline
\end{tabular}
}
\label{tab2}
\end{center}
\end{table*}

\begin{table*}[!h]
\centering
\caption{Data collected from Java Project 2}
\begin{center}
\resizebox{\linewidth}{!} {
\begin{tabular}{|c|c|c|c|c|c|c|c|c|c|c|c|}
\hline
\multicolumn{7}{|c|}{\textbf{Source Monitor On Java Project 2}}& \multicolumn{5}{|c|}{\textbf{PMD on Java Project 2}} \\
\hline
Commit & Type & \%Branches & Class & Methods/Class & Avg Statement/Method & Avg Depth & Avg Complexity & Best Practices & Code Style & Design Documentation & Error Prone \\
\hline
Before & 5.1 & 52 & 2.69 & 2.57 & 1.15 & 1.50 & 42 & 586 & 178 & 266 & 134 \\
\hline
After & 5.6 & 57 & 2.95 & 2.61 & 1.22 & 1.62 & 44 & 676 & 280 & 314 & 158\\
\hline
Before & 7.2 & 134 & 3.63  & 2.66 & 1.44 & 1.58 & 31 & 457 & 154 & 189 & 98 \\
\hline
After & 7.4 & 135 & 3.61 & 2.67 & 1.44 & 1.58 & 31 & 449 & 154 & 189 & 98\\
\hline
Before & 7 & 62 & 3.24 & 2.96 & 1.33 & 1.62 & 46 & 663 & 204 & 279 & 135\\
\hline
After & 6.3 & 63 & 3.19 & 2.85 & 1.3 & 1.56 & 46 & 643 & 204 & 279 & 135\\ 
\hline
Before & 6.2 & 95 & 3.08 & 2.72 & 1.29 & 1.54 & 56 & 785 & 277 & 329 & 157\\
\hline 
After & 6.2 & 95 & 3.08 & 2.72 & 1.29 & 1.54 & 46 & 643 & 204 & 279 & 135\\
\hline
Before & 6.9 & 100 & 3.79 & 2.74 & 1.49 & 1.50 & 56 & 1852 & 564 & 882 & 328 \\
\hline
After & 6.2 & 95 & 3.08 & 2.72 & 1.29 & 1.54 & 56 & 1846 & 560 & 1006 & 324 \\
\hline 
Before & 6.1 & 126 & 3.71 & 2.4 & 1.42 & 1.41 & 58 & 1846 & 560 & 1014 & 324 \\
\hline 
After & 6.1 & 126 & 3.67 & 2.41 & 1.41 & 1.41 & 18 & 1786 & 530 & 994 & 324 \\
\hline Before & 6.1 & 126 & 3.67 & 2.41 & 1.42 & 1.41 & 18 & 2008 & 654 & 1186 & 382 \\
\hline After & 5.9 & 146 & 3.61 & 2.4 & 1.41 & 1.43 & 18 & 2066 & 674 & 1210 & 388 \\
\hline
Before & 7.3 & 156 & 3.46 & 2.65 & 1.39 & 1.57 & 74 & 2280 &722 &1170 & 428\\
\hline
After & 6.7 & 159 & 3.4 & 2.66 & 1.24 & 1.58 & 66 & 2434 & 700 & 1128 & 454\\
\hline
Before & 6.4 & 149 & 3.6 & 2.63 & 1.25 & 1.60 & 58 & 2372 & 692 & 1164 & 480\\
\hline
After & 6.4 & 153 & 3.44 & 2.62 & 1.19 & 1.56 & 50 & 2324 & 714 & 1124 & 482\\

\hline

\end{tabular}
}
\label{tab3}
\end{center}
\end{table*}

\subsection{Additional data collected from Source monitor and PMD:}\label{SCM}

We got some other data from Source Monitor and PMD after running on the
selected versions/states(before and after refactoring) of projects. 

\begin{table*}[!h]
\centering
\caption{Data collected from Java Project 1}
\begin{center}
\resizebox{\linewidth}{!} {
\begin{tabular}{|c|c|c|c|c|c|c|c|c|}
\hline
\multicolumn{8}{|c|}{\textbf{Source Monitor On Java Project 1}}& \multicolumn{1}{|c|}{\textbf{PMD On Java Project 1}} \\
\hline
Commit Type & Files & Lines & Statements & Calls(Methods) &
\%Comments & Max Complexity & Max Depth & Multithreading \\
\hline
Before & 20 & 1200 & 721 & 236 & 7.1 & 7 & 4  & 3 \\
\hline
After & 22 & 1373 & 800 & 267 & 5.7 & 7 & 5 & 3\\
\hline
Before & 26 & 1771 & 1027 & 333 & 8.9 & 7 & 5 & 4\\
\hline
After & 26 & 1776 & 1026 & 333 & 9.2 & 7 & 5 & 4 \\
\hline
Before & 31 & 2214 & 1258 & 451 & 7.9 & 11 & 5 & 4 \\
\hline
After & 32 & 2247 & 1281 & 464 & 7.8 & 11 & 5 & 4 \\
\hline
Before & 32 & 2247 & 1281 & 464 & 7.8 & 11 & 5 & 3\\
\hline
After & 38 & 2712 & 1647 & 633 & 6.8 & 13 & 7 & 3 \\
\hline
Before & 118 & 6535 & 4314 & 1328 & 9 & 15 & 8 & 16 \\
\hline 
After & 118 & 6526 & 4306 & 1328 & 8.9 & 15 & 8 & 16\\
\hline Before & 103 & 5952 & 3868 & 1262 & 9.5 & 15 & 8 & 16\\
\hline After & 103 & 5925 & 3851 & 1259 & 9.4 & 15 & 8 & 16 \\
\hline Before & 100 & 5900 & 3830 & 1259 & 9.5 & 15 & 8 & 9\\
\hline After & 93 & 5760 & 3723 & 1249 & 9.5 & 15 & 8 & 9\\
\hline Before  & 73 & 5336 & 3388 & 1192 & 10.2 & 15 & 8 & 16\\
\hline
After & 73 & 5218 & 3314 & 1161 & 10.3 & 15 & 8 & 16\\
\hline

\end{tabular}
}
\label{tab4}
\end{center}
\end{table*}

\begin{table*}[!h]
\centering
\caption{Data collected from Java Project 2}
\begin{center}
\resizebox{\linewidth}{!} {
\begin{tabular}{|c|c|c|c|c|c|c|c|c|}
\hline
\multicolumn{8}{|c|}{\textbf{Source Monitor On Java Project 2}}& \multicolumn{1}{|c|}{\textbf{PMD On Java Project 2}} \\
\hline
Commit Type & Files & Lines & Statements & Calls(Methods) &
\%Comments & Max Complexity & Max Depth & Multithreading \\
\hline
Before & 47 & 1545 & 991 & 249 & 20.5 & 6 & 4 & 4 \\
\hline
After & 51 & 1782 & 1157 & 299 & 19.1 & 12 & 4 & 4\\
\hline
Before & 126 & 5545 & 3144  & 831 & 29.5 & 11 & 7 & 2\\
\hline
After & 127 & 5576 & 3181 & 838 & 29 & 11 & 7 & 2 \\
\hline
Before & 56 & 2124 & 1417 & 401 & 17.7 & 10 & 5 & 5\\
\hline
After & 57 & 2109 & 1406 & 401 & 18.1 & 10 & 5 & 5\\
\hline

Before & 83 & 3008* & 1984 & 545 & 18.4 & 11 & 5  & 6 \\
\hline
After & 83 & 3008 & 1984 & 545 & 18.4  & 11 & 5 & 5 \\
\hline 
Before & 89 & 3614 & 2390 & 714 & 17 & 9 & 7 & 146\\
\hline After & 83 & 3008 & 1984 & 545 & 18.4 & 11 & 5  & 14\\
\hline Before & 120 & 4,360 & 2704 & 745 & 21.6 & 8 & 7 & 14 \\
\hline After &120 & 4,668 & 2695 & 743 & 27 & 8 & 7 & 14\\
\hline Before & 120 & 4,667 & 2694 & 741 & 27 & 8 & 7 & 14\\
\hline After  & 140 & 5,434 & 3145 & 816 & 26.6 & 10 & 7 & 14\\

\hline

\end{tabular}
}
\label{tab5}
\end{center}
\end{table*}

\section{Result}

\subsection*{\textbf{RQ1: Do refactoring commits have any impact on quality metrics? }}
We applied Source Monitor, and PMD on the java projects and Sonarqube on the python projects. But before that, we had to detect refactoring commits by running Refactoring Minor 2.1 and running the static analysis tools(Source Monitor, PMD, Sonarqube ) on the selected pairs manually. Selected pairs are the versions of the project before and after refactoring commits.

After running the Source Monitor, we got some attributes. Some of them are method/class and average statements/methods. Software quality is inversely proportional with method/class as a class with fewer methods is easy to maintain and Single Responsibility Principle is preserved. Software quality is inversely proportional to method/class. That means if the value of the method/class is low then the quality of the software is better.

From \tablename~\ref{tab2}  and \tablename~\ref{tab3} , we can see that the value of the method/class is changing. Most of the time the value is changing positively(decreasing the value of the method/class). But there are some unusual cases also where the value is changing in the reverse order. And this is also true for average statements/methods. 

From \tablename~\ref{tab2}  and \tablename~\ref{tab3}, we can also see that average depth and average complexity is also changing after the refactoring commit. If the average depth increases, there is a possibility of increasing dependency which may result in coupling. And all of these are increasing in complexity. It is hard to maintain any complex class. If the complexity increases, it creates problems in maintainability. Things get complicated in analyzing, testing, and changing any module. It is hard to understand any complex module.

\begin{figure*}[hbtp]
    \centering
    {\includegraphics[width=\textwidth]{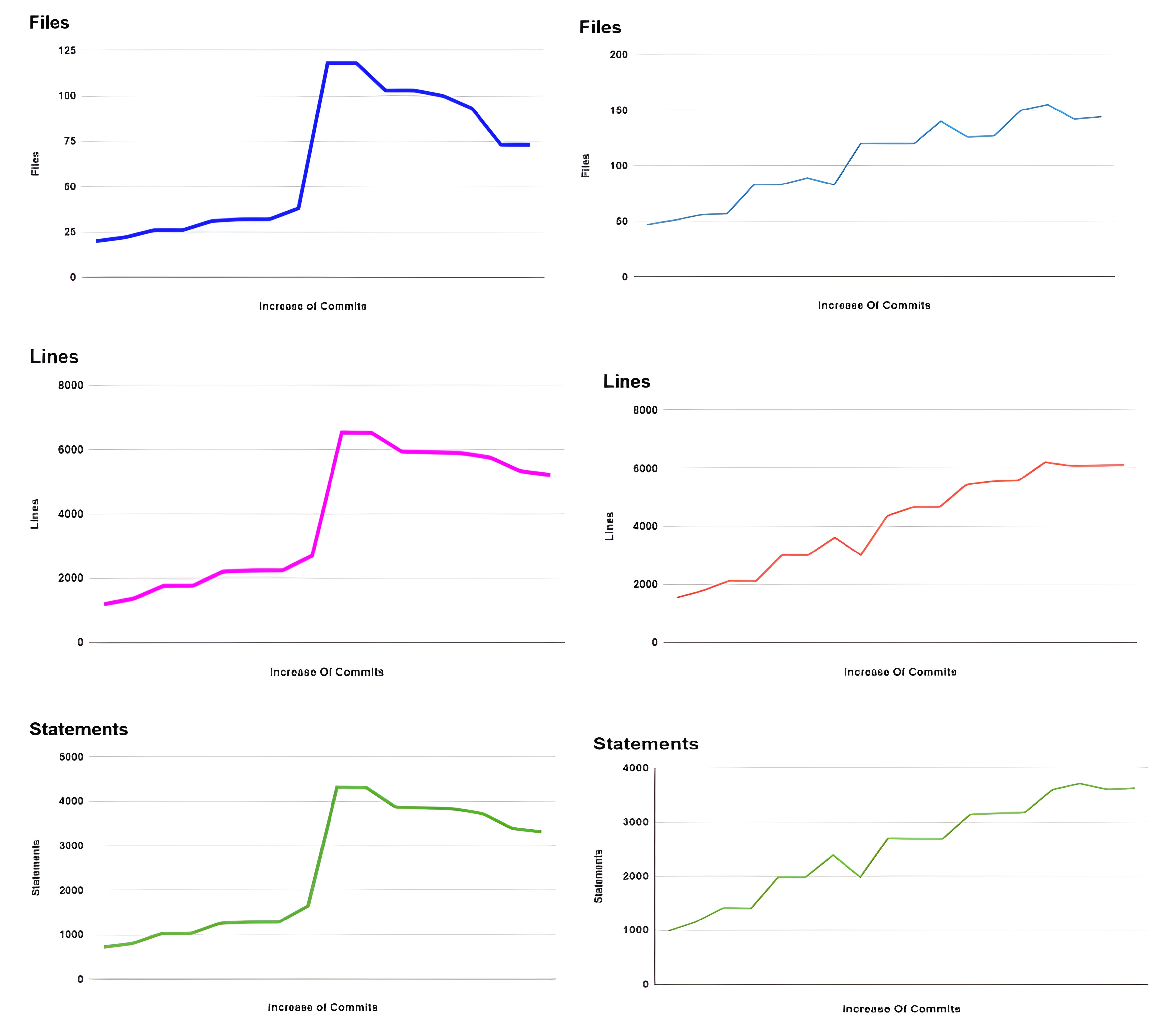}}
    \caption{Impact of Refactoring Commit on Size (File, Line, Statement) in Java Project 1(column 1) and Java Project 2(column 2)}
        \label{fig:1}
\end{figure*}

It is visible that after refactoring, generally the size of the software increases. This size is represented by the file, size, and statement in \tablename~\ref{tab4}  and \tablename~\ref{tab5}. We can also visualize this in Figure \ref{fig:1}.

\subsection*{\textbf{RQ2: What is the impact of refactoring on code smells?}}
To examine the impact of refactoring on code smell, we analyzed the data obtained from a total of 62 revision files, 13 pairs from python project 1 and 18 pairs from python project 2. We observed 3 types of change –

\begin{itemize}
    \item Positive change where the no of code smells reduced after refactoring commit
    \item Negative change where the no of code smells increased after refactor commit and
    \item  Unchanged where the no of code smells remained the same after refactoring commit.
    
\end{itemize}

Therefore, in python project 1, we detected 46\% positive impact, 15.3\% negative impact, and 38\% neutral impact of refactoring. In python project 2, we observed the positive impact on 9 pairs and the negative impact on 3 pairs. In the remaining 6 pairs, the impact was unchanged. We detected a 50\% positive impact, 16.67\% negative impact and 33.33 \% neutral impact of refactoring.

\begin{itemize}
    \item Impact = no of (positive/negative/unchanged) pair / Total no of pair * 100 \%
\end{itemize}

\begin{figure*}[hbtp]
    \centering
    {\includegraphics[width=\textwidth]{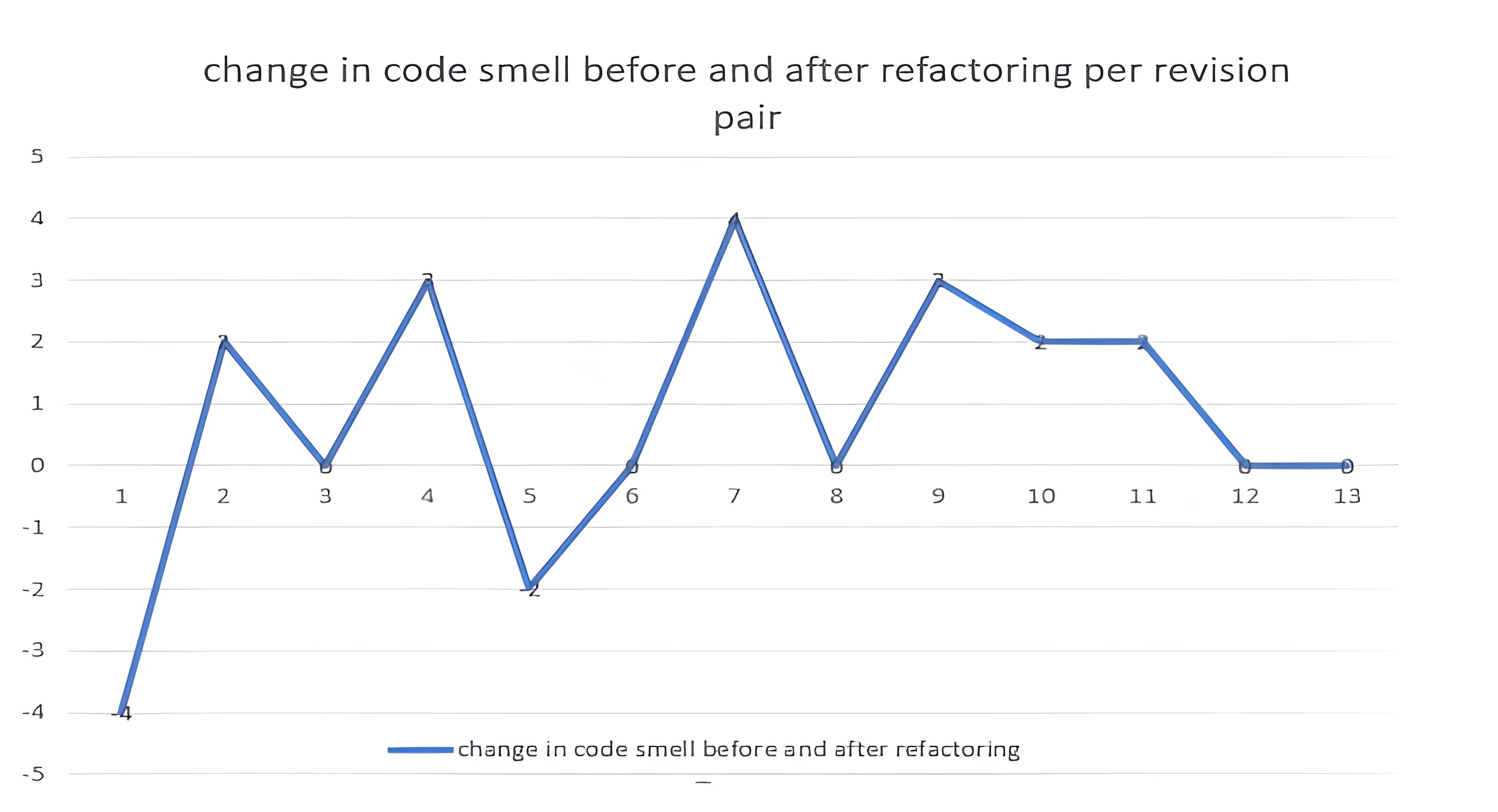}}
    \caption{ Difference Value Of Code smells Before and After Refactoring(Python Project 1)}
    \label{fig:4}
\end{figure*}

\begin{figure*}[hbtp]
    \centering
    {\includegraphics[width=\textwidth]{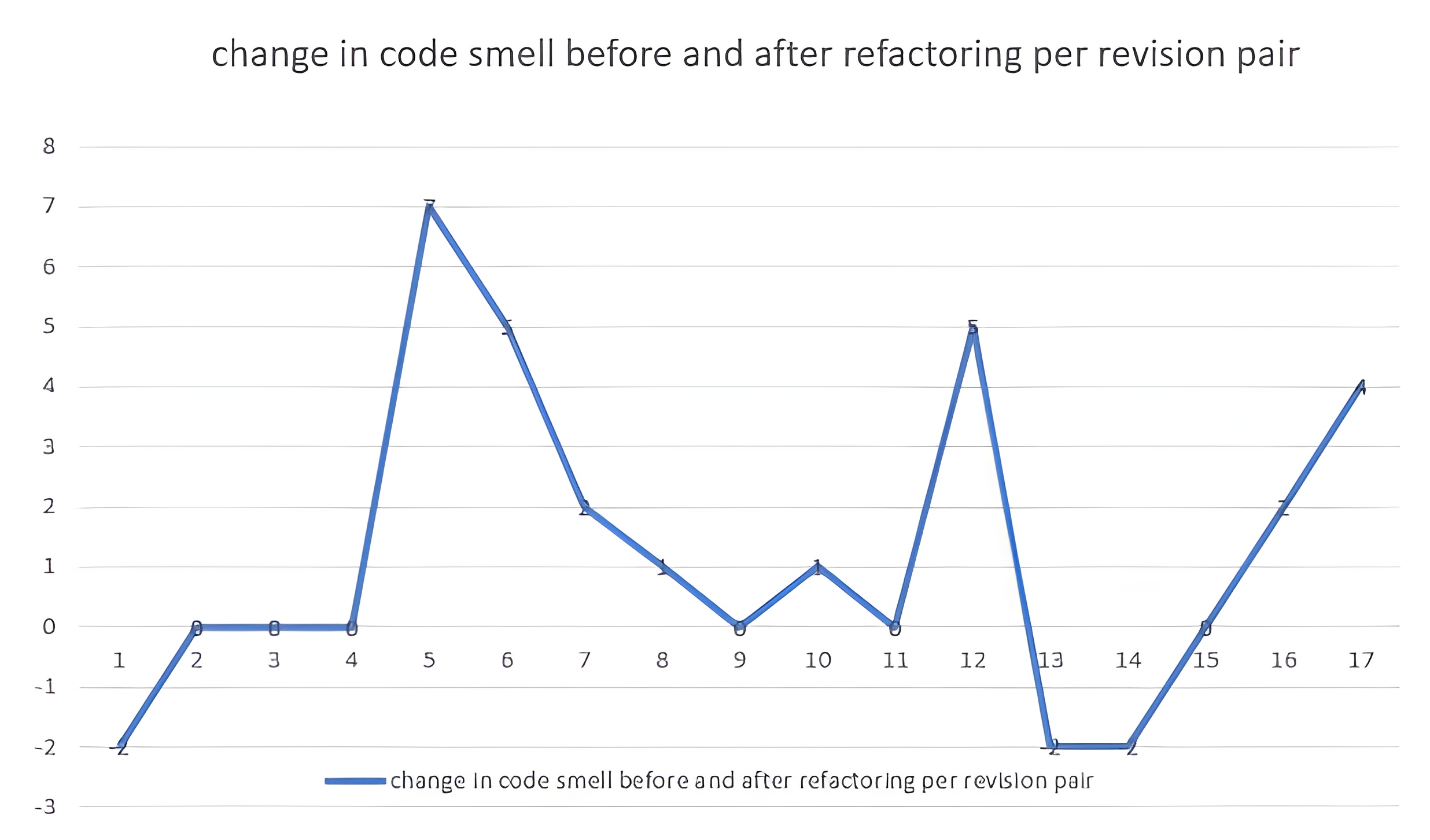}}
    \caption{Difference Value Of Code smells Before and After Refactoring(Python Project 2)}
    \label{fig:5}
\end{figure*}

We also calculated the value difference of code smell between the refactoring commit and the commit before that. If the no of code smell is reduced after refactoring commit, it indicates a positive change. And if increased, the change is negative. The result is presented in Figure \ref{fig:4} for python project 1 and Figure \ref{fig:5}  for python project 2. 

Value Difference of code smell = no of code smell before refactoring – no of code smell after refactoring

\begin{figure}[htbp]
\centerline{\includegraphics{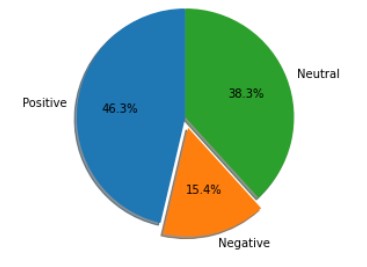}}
\caption{ Impact on python project 1}
\label{fig:2}
\end{figure}

\begin{figure}[htbp]
\centerline{\includegraphics{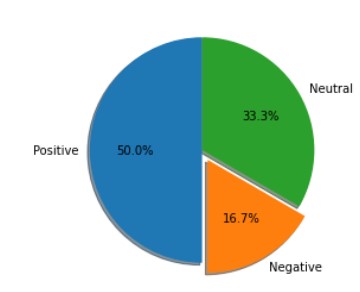}}
\caption{ Impact on python project 2}
\label{fig:3}
\end{figure}

We also observed that in the first project, 7\% refactoring was committed and in python project 2, the value is 17.7\%. The number of code smells tends to be higher in projects where less refactoring is applied. In both projects, we also found the value of debt. Debt is the technical effort to fix the code smell. Debt is proportional to code smell, it indicates that the more code smell found, the more time and effort would be required to fix the smell. In each refactor commits, developers are either introducing new code smells or resolving the code smells, or making no impact. So it is necessary to utilize proper tools to detect and realize new smells. Otherwise, they will need to dedicate more time and effort to maintaining the system.

\subsection*{\textbf{RQ3: Does the refactoring commit impact the same way on novice and mature projects?}}

A total of four projects were taken for our experiment. Two of them were java projects and two were Python projects. We had the intention to find out whether these refactoring commits impacted the same way on novice and mature projects. So we chose projects in a way that we can distinguish the results of these two types. We chose three mature projects and one novice project. When choosing mature projects, we considered a few points. We considered projects with more than 250 stars, forks, and tags as expert projects. On the other hand, we chose a project where the value of all these features is near zero, and we considered that project a novice one.
We took 11 quality metrics in total to calculate the impact of the two Java projects. Commits are taken pairwise, where they were selected as a sequence of before and after refactoring commits. We address RQ2 by analyzing the behavior of quality metrics in three different ways: positively, negatively, and neutrally. We set priority values for all the quality metrics. If a pair positively impacted on a metric, then we assigned the value 1 for that metric; if negative, then -1; otherwise, 0 for neutral impact. Then we sum the values for all metrics. If the summation gave us a positive value (more than 0), then we assigned the commit pair as a positive pair, negative for less than 0 value, and neutral for those pairs whose summation of value was 0. In this way, we calculated values for all commit pairs.

\begin{figure}[htbp]
\centerline{\includegraphics{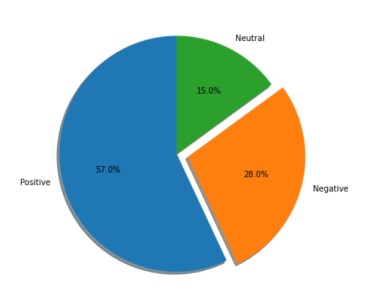}}
\caption{Impacts On Novice Projects}
\label{fig:6}
\end{figure}

\begin{figure}[htbp]
\centerline{\includegraphics{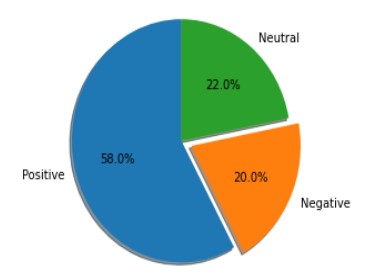}}
\caption{Impacts On Mature Projects}
\label{fig:7}
\end{figure}

We had One project developed by novice developers and 3 projects developed by expert developers. For analyzing the result we took the direct value from the novice project and the avg of 3 values for the expert projects. From the pie charts \ref{fig:2}, \ref{fig:3}, we can visualize that the positive impact of refactoring commits on both two types of projects was quite the same. Expert projects stand slightly higher than novice projects. But the difference is clearly noticeable for neutral and negative impacts. Projects developed by expert developers have less negative and more neutral impacts than projects developed by novice developers.

\subsection*{\textbf{RQ4: How does refactoring commit impact projects over time?}}

\begin{figure}[htbp]
\centerline{\includegraphics{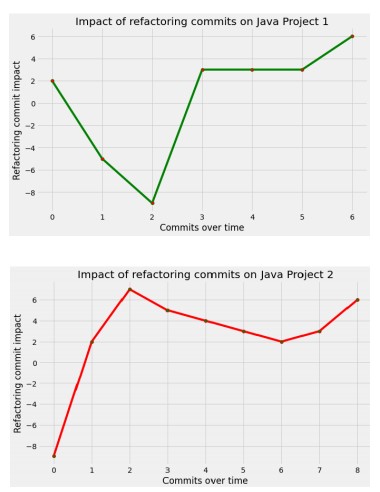}}
\caption{Impact of Refactoring Commits Over Time.}
\label{fig:8}
\end{figure}

We took all pairs of commits modified by the date. Our goal was to find whether these refactoring commits impacted the same way on the projects over the period or not. As we know, the overall condition of the projects does not remain the same throughout the whole development phase. As we mentioned earlier, we address commit pairs by analyzing the behavior of quality metrics in three different approaches: positive, negative, and neutral. With these values, we plot graph \ref{fig:8} for two java projects to find the impact of refactoring commits over time.

\section{Limitations}

\subsection*{Internal validity:} We mostly relied on static code analysis and mining tools. We also set the default configuration of the tools. So the accuracy of the data collection procedure and analysis is dependent on the tool results. However, we selected some popular tools widely used by developers. Moreover, the average precision of Refactoring Miner equals 96.4\% with low rates of false positives. 
\subsection*{External validity:} The result of our study is based on four selected open-source systems as it is more convenient to access those projects for us. So it is difficult to generalize to a larger context. 
\subsection*{Conclusion validity:} Due to time constraints and the depth of our selected study, we are bound to analyze a small data sample. In addition, we have selected the commit before refactoring manually, so there might be a chance of human error.

\section{Conclusion}

We performed an empirical study on four open-source software projects to understand the impact of refactoring and its relation to the evolution of software quality. We ran different static code analyzer tools on the projects of python and java separately. The tools give some value to software quality metrics from which we try to make a study on software quality. 

We have observed three types of impact on refactoring - positive, negative, and neutral; by calculating the number of code smells before and after a refactoring operation is committed on GitHub by the developers. After analyzing the data obtained from python projects, we observed that refactoring has a mostly positive impact( 46\% and 50\% in two Python projects respectively). Refactoring operations also have the chance to introduce new code smells if not performed tactfully, which we observed as a negative impact (15.3\% and 16.6\% in two python projects respectively). We concluded that the number of code smells is directly proportional to the technical effort to fix it. We also noticed that the same number of code smells need less effort to fix if identified at an earlier stage. Moreover, between the 2 projects, the one where refactoring was applied frequently tended to have less code smell than the other. 

We examined the impact of refactoring commits by detecting the change in quality attributes by executing static analysis tools on two Java systems as well. Analysis of the results of two static analysis tools on those java projects indicates that refactoring impacts much on code quality metrics. Mostly they were positive impacts(57\% and 70\% in two java projects respectively) but some negative impacts(28\% and 30\% in two java projects respectively) had also been found. In java project 1, some(15\%) of impacts were neutral. It is noticeable that the positive impact of refactoring increases at the middle to later stages of the development process. 

From this analysis, we can conclude that Refactoring commits identically impacts on quality metrics. Though sometimes it may affect negatively, most of the time it affects positively. The positive effect increases with the evaluation of the development process. In the future, we want to increase the amount of data collection. We will increase the number of projects, refactoring commits to get a more accurate value. We will include projects of different languages, and static code analyzers to get more software quality metrics. We'll also use dynamic analysis tools and regression testing to figure out what else changes in the program and why it changes.

\section*{Acknowledgment}

I would like to express my deepest gratitude to everyone who contributed to this paper.

I would like to thank my supervisor for his direction and assistance throughout the entire research process. His insightful observations and suggestions were crucial in shaping the direction of this paper.

My family and friends, who have always been a source of inspiration and motivation for me, also deserve my deepest appreciation. Their unwavering support and encouragement were crucial in assisting me in overcoming the numerous obstacles encountered during the research process.

Once more, I would like to extend my sincere appreciation to everyone who has helped to complete this paper.






\vspace{12pt}

\end{document}